\documentclass[onecolumn,10pt]{article} 

\usepackage[square,numbers,sort&compress,comma]{natbib}

\usepackage{amsmath}
\usepackage{amssymb}
\usepackage{caption}
\usepackage{graphicx}
\usepackage{latexsym}
\usepackage{times}
\usepackage[pagewise]{lineno}
\usepackage{bm}
\usepackage{lineno,hyperref}

\topmargin - 12pt 
\renewenvironment{abstract}%
              {
               \small
               {\bfseries \abstractname}
               \par
               \vspace{10pt}
              }

\renewcommand\abstractname{Abstract}

\newcommand{\nomenclature}
              [1]
              {
               \bgroup
               \flushleft
               \small\bf
               #1
               \par
               \egroup
              }

\renewcommand{\section}
              [1]
              {
               \bgroup
               \flushleft
               \small\bf
               \refstepcounter{section}
               \arabic{section}. #1
               \par
               \egroup
              }

\renewcommand{\subsection}
              [1]
              {
               \bgroup
               \flushleft
               \small\em
               \refstepcounter{subsection}
               \arabic{section}.
               \arabic{subsection}. #1
               \par
               \egroup
              }

\renewcommand{\subsubsection}
              [1]
              {
               \bgroup
               \flushleft
               \small\em
               \refstepcounter{subsubsection}
               \arabic{section}.
               \arabic{subsection}.
               \arabic{subsubsection}. #1
               \par
               \egroup
              }

  \newcommand{\acknowledgement}
              [1]
              {
               \bgroup
               \flushleft
               \small\bf
               #1
               \par
               \egroup
              }

  \newcommand{\sectionbib}
              [1]
              {
               \bgroup
               \flushleft
               \small\bf
               #1
               \par
               \egroup
              }

\begin{document}

\title{\LARGE Evaluation of flamelet-based models for liquid ammonia combustion in a temporally evolving mixing layer}

\author{{\large Zhenhua An, Jiangkuan Xing$^{*}$, Abhishek Lakshman Pillai, Ryoichi Kurose}\\[10pt]
        {\footnotesize \em Department of Mechanical Engineering and Science, Kyoto University, Kyoto, daigaku-Katsura, Nishikyo-ku, Kyoto}\\[-5pt]
        {\footnotesize \em 615–8540, Japan}\\[-5pt]
       }

\date{}


\small
\baselineskip 10pt


\vspace{50pt}
\maketitle
\noindent \rule{\textwidth}{0.5pt}

\begin{abstract} 
Liquid ammonia combustion can be enhanced by co-firing with small molecular fuels such as methane, and liquid ammonia will undergo flash evaporation due to its relatively low saturation pressure. These characteristics, involving the presence of multiple fuel streams, a rapid phase change process, and strong heat loss, pose challenges for flamelet modeling of liquid ammonia combustion. To address these issues, this study aims to evaluate the effectiveness of flamelet-based models for liquid ammonia combustion in a turbulent mixing layer. Specifically, the extended flamelet/progress variable (E-FPV), extended flamelet-generated manifolds (E-FGM), and extended hybrid (E-Hybrid) models are developed and assessed. Firstly, a three-dimensional Point-Particle Direct Numerical Simulation (PP-DNS) with detailed chemistry is performed, where the turbulent flow is fully resolved, and the ammonia droplets are described by the Lagrangian method, to investigate the combustion characteristics of a liquid ammonia/methane co-fired flame and to provide state-of-the-art validation data for flamelet modeling. The PP-DNS results reveal distinct stages in the liquid ammonia/methane co-fired flame, namely, the mixing, methane-dominated, and fully reacting stages. The phase change process introduces significant heat loss due to the high latent heat of liquid ammonia. Subsequently, flamelet-based models are developed to account for the complex fuel streams, rapid phase change process, and strong local heat loss. The performance of these models is evaluated through \textit{a priori} analysis by comparing the predictions with the PP-DNS results. The \textit{a priori} results show that the E-FGM model outperforms the E-FPV and E-Hybrid models. This superior performance can be attributed to the rapid flash evaporation and sufficient mixing of the superheated ammonia, resulting in the dominance of the premixed combustion mode in liquid ammonia combustion.
\end{abstract}

\vspace{10pt}
\parbox{1.0\textwidth}{\footnotesize {\em Keywords:} Liquid ammonia combustion; Point-Particle DNS; FPV; FGM; Hybrid model; Turbulent mixing layer}
\rule{\textwidth}{0.5pt}
\vspace{10pt}

*Corresponding author.\\
\textit{E-mail address:} xing.jiangkuan.6h@kyoto-u.ac.jp (JK. Xing).

\clearpage

\newpage

\section{Introduction\label{sec:introduction}} \addvspace{10pt}

As a hydrogen carrier and a renewable carbon-free fuel, ammonia plays a critical role in the global energy supply chain, facilitating the achievement of decarbonization goals. Over the past few decades, significant progress has been made in ammonia combustion research \cite{valera2018ammonia, kobayashi2019science}. In recent years, there has been a notable increase in research focused specifically on the combustion of liquid ammonia, emerging as a new focus. Compared to gaseous ammonia, the liquid ammonia combustion offers several advantages. It simplifies the fuel supply and pre-evaporations systems due to the liquid storage form of ammonia. Notably, liquid ammonia has an exceptional octane number exceeding 110, making it suitable for internal combustion engines \cite{li2022investigation}. In addition, the use of liquid ammonia can effectively reduce energy consumption during the vaporization process and minimize start-up time for gas turbines \cite{okafor2021flame, okafor2021liquid}.

To date, some research have been conducted to investigate the spray and combustion characteristics of liquid ammonia. Okafor et al. \cite{okafor2021flame, okafor2021liquid, somarathne2023towards} conducted experiments and simulations to examine the direct combustion of liquid ammonia spray in a swirl combustor, focusing on stability and emission characteristics. Mounaïm-Rousselle et al. \cite{pele2021first, pandal2023gdi, pele2023spray, zembi2023numerical} systematically conducted the research on liquid ammonia spray characteristics and evaporation process through experiments and simulations. Li et al. \cite{zhang2021injection, li2022investigation, li2022comparison, zhou2023pilot, yi2023simulation} systematically investigated the liquid ammonia combustion under internal combustion engine conditions, covering aspects ranging from spray injection characteristics to the combustion of diesel-ignited ammonia dual fuel. In addition, Colson et al. \cite{colson2023study} studied the effect of injection temperature and nozzle geometry on ammonia flash spray. Scharl et al. \cite{scharl2023characterization} investigated the ammonia spray combustion and mixture formation under internal engine-like conditions. The above summarizes the majority of research conducted on the spray and combustion of liquid ammonia, relying primarily on experimental investigations. However, numerical simulations of liquid ammonia remain limited, especially in the area of combustion simulation. Regarding spray simulation, An et al. \cite {an2023numerical} developed a combined phase change model for liquid ammonia. Meanwhile, Zhang et al. \cite{zhang2023numerical}, Shin et al. \cite{shin2023ammonia}, Zembi et al. \cite{zembi2023numerical}, and Huang et al. \cite{huang2023large} also evaluated the phase change models and investigated the special characteristics of ammonia flash spray. In terms of combustion simulation, An et al. \cite{an2024numerical} conducted a numerical study of spherical flame propagation in dispersed liquid ammonia droplets using the quasi-Direct Numerical Simulation (DNS) method. Somarathne et al. \cite{somarathne2023towards} numerically studied the liquid ammonia/air spray combustion in a gas turbine-like combustor using Large Eddy Simulation/partially stirred reactor (LES/PaSR) approach.

The flamelet model has attracted a lot of attention in combustion simulations due to its ability to reduce computational cost while considering detailed chemistry \cite{hu2019partially}. However, to the best of the authors’ knowledge, there is currently no relevant research on the simulation of liquid ammonia flames using flamelet models. The unique flash evaporation of liquid ammonia leads to rapid phase change and significant local heat loss. In the case of liquid ammonia flames, stabilization often requires the assistance of pilot flames fueled by small-molecule fuels, resulting in the presence of multiple fuel streams. The applicability of flamelet-based combustion models to liquid ammonia combustion and how to account for the presence of multiple fuel streams and strong heat loss in liquid ammonia flames remain open questions.

To this end, the present study aims to assess the performance of flamelet-based models for liquid ammonia combustion modeling. First, a three-dimensional Point-Particle Direct Numerical Simulation (PP-DNS) is performed to investigate the combustion characteristics of a liquid ammonia/methane co-fired flame in a temporally evolving mixing layer. This PP-DNS simulation also serves as a reference for evaluating the subsequent flamelet models. Second, three flamelet models, including the extended flamelet progress variable model (E-FPV), the extended flamelet generated manifold model (E-FGM), and the extended hybrid model (E-Hybrid) that combines the FPV and FGM through flame index, are developed and examined through the \textit{a priori} analysis with the PP-DNS solutions as benchmarks.

\section{Mathematical approaches\label{sec:sections}} \addvspace{10pt}

\subsection{PP-DNS coupled with detailed chemistry\label{subsec:subsection}} \addvspace{10pt}

In the PP-DNS, the gas phase governing equations, including the continuous, momentum, enthalpy, and species equations, are solved without any averaging and filtering. The detailed chemistry (59 species and 356 elementary reactions \cite{okafor2018experimental}) and detailed molecular mixing are considered. For the dispersed phase, the conservation equations, including the position, velocity, and temperature are solved under the Lagrangian framework. The particular phase change process of ammonia droplets is described using a combined model developed and validated in our previous study \cite{an2023numerical}. The two-way coupling method is used to describe the interactions between the gas phase and the droplets. The details of all conservation equations are not presented here for brevity and the interested reader can refer to our previous studies \cite{an2024numerical, xing2023extended}.

\subsection{Extended flamelet-based models\label{subsec:subsection2}} \addvspace{10pt}

Flamelet-based models assume that the thermo-chemical state of the flame can be tabulated and expressed as a function of several control variables. In general, diffusion and premixed flames can be described by FPV \cite{pierce2004progress} and FGM \cite{van2016state} models, respectively. In the present work, three models are developed, named the E-FPV, E-FGM, and E-Hybrid models. Compared with the traditional FPV and FGM models, we introduced two additional parameters, $X$ (methane ratio in the whole fuel stream) and $h$ (total enthalpy), to consider the complex fuel streams and heat loss. Therefore, 4 control variables, i.e., progress variable ($c$), mixture fraction ($Z$), methane ratio ($X$), and total enthalpy ($h$), are used to describe the liquid ammonia/methane co-fired flame. Specifically, the thermochemical properties $\varphi$ can be tabulated as: 
\begin{equation}
	\varphi = \varphi \left(c, Z, X, h_{norm} \right) ,
\end{equation}
where $c$ is defined as the sum of the mass fractions of H$_2$O, CO$_2$, and H$_2$ obtained directly from the DNS results in the \textit{a priori} study. $Z$ can be calculated as $Z = Z_{\mathrm{CH_4}} + Z_{\mathrm{NH_3}}$. $Z_{\mathrm{CH_4}}$, $Z_{\mathrm{NH_3}}$, and $Z$ are the methane, ammonia, and total mixture fractions, respectively. The transportation equations of $Z_{\mathrm{CH_4}}$ and $Z_{\mathrm{NH_3}}$ are also solved in the DNS. $X = Z_{\mathrm{CH_4}}/\left(Z_{\mathrm{CH_4}} + Z_{\mathrm{NH_3}}\right)$. $h$ is normalized during the table lookup process.

In the flamelet-based models, the flow field is described by the conservation equations of mass and momentum. Additional transport equations of progress variable, mixture fractions, and total enthalpy are solved as:
\begin{equation}
	\frac{\partial\left(\rho {c} \right)}{\partial t}+\nabla \cdot \left( \rho {\bm u} {c}\right) = \nabla \cdot \left(\rho D_{c} \nabla {c}\right)+\dot{\omega}_c+\dot{S}_c,
\end{equation}
\begin{equation}
	\frac{\partial\left(\rho {Z_{i}} \right)}{\partial t}+\nabla \cdot \left( \rho {\bm u} {Z_{i}}\right) = \nabla \cdot \left(\rho D_{i} \nabla {Z_i}\right)+\dot{S}_i ,
\end{equation}
\begin{equation}
	\frac{\partial \left(\rho  h \right)}{\partial t} + \nabla\cdot\left(\rho {\bm u}  h\right) = \nabla\cdot \left(  \rho D_h \nabla h\right) + {\dot S_h} ,
\end{equation}
where $\rho$ and $\bm u$ are the gaseous phase density and velocity, respectively. $D_k$ represents the diffusion coefficient of $k$. $\dot{\omega}_c$ is the reaction rate derived from the flamelet table. $Z_i$ represents the $Z_{\mathrm{CH_4}}$ or $Z_{\mathrm{NH_3}}$, and $\dot{S}_i$ equals to 0 when $i = _\mathrm{CH_4}$.

To construct the flamelet tables, one-dimensional counter-flow diffusion flames and freely-propagating flames are calculated under different $X$ values and initial temperatures. All the flamelet tables are obtained using the FlameMaster package \cite{pitsch1998flamemaster} with the same reaction mechanism as in the DNS (59 species and 356 elementary reactions \cite{okafor2018experimental}). $X$ ranges from 0 to 1 with an interval of 0.1. When $X$ value equals to 0 or 1, the local fuel would be pure ammonia or methane, respectively. For E-FPV tables, one-dimensional counter-flow diffusion flames are calculated at the different stoichiometric scalar dissipation rate from 0.001 s$^{-1}$ to a maximum value where the flame is completely quenched. The fuel temperature $T_f$ is set to be equal to the oxidizer temperature $T_o$ to represent the interphase heat transfer and ranges from 250 to 2100 K. Note that the 250 K is used here to account for the large heat loss caused by flash evaporation. For E-FGM tables, one-dimensional freely-propagating flames are calculated at different equivalence ratios, which range from 0.4 to 2.0 with an interval of 0.1. The flamelets under the different equivalence ratios are then converted to $Z$ space. Different initial temperatures from 250 to 700 K are considered to represent the heat loss effect. In addition, pure oxidizer and fuel solutions are also created to ensure the completeness of the E-FGM tables in the mixture fraction space. In the present work, the E-FPV and E-FGM flamelet tables are generated with a resolution of $150\times51\times11\times51$ for $c \times Z \times X \times h_{norm}$, respectively. For the E-Hybrid model, the flame index which is defined as \cite{yamashita1996numerical} to access the local combustion mode, $FI = 1/2 \left(1+\left(\nabla Y_F \cdot \nabla Y_{O_2}\right)\left(\left|\nabla Y_F \cdot \nabla Y_{O_2} \right|+\epsilon\right)\right)$, where $Y_F$ and $Y_{O_2}$ are the mass fractions of fuel and oxygen, respectively. $\epsilon$ is an infinitesimal positive value to avoid dividing by zero. The final prediction of the E-Hybrid model is obtained by $\varphi_{E-Hybrid} = FI \varphi_{E-FGM}+\left(1-FI\right) \varphi_{E-FPV}$.

\begin{figure}[h!]	
	\centering
	\includegraphics[width=192pt]{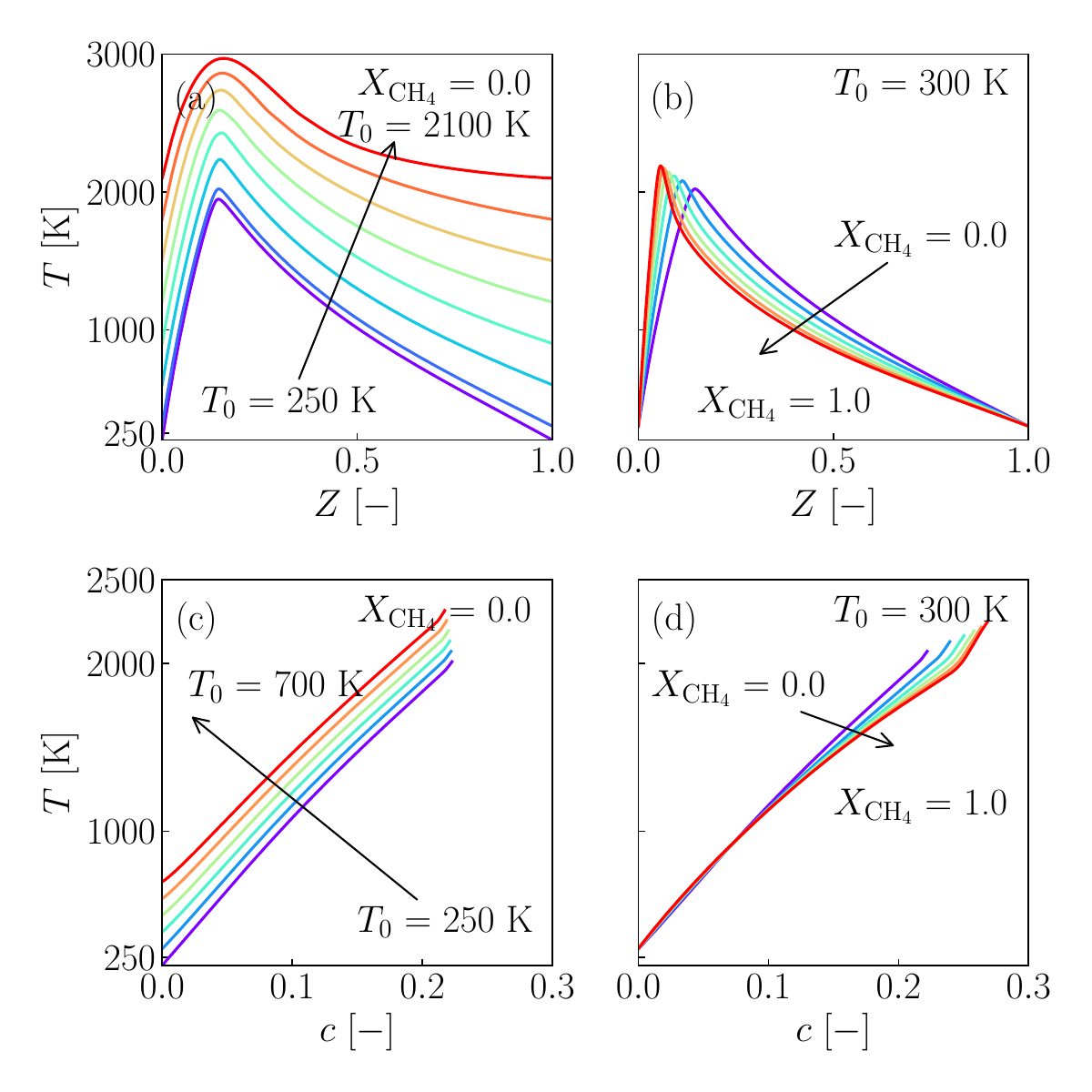}
	\caption{\footnotesize 1-D flamelet profiles of temperature against the $Z$ and $c$. (a) FPV flamelet data at $X_{\mathrm{CH4}}$ = 0.0, $T_0$ = 250 - 2100 K. (b) FPV flamelet data at $T_0$ = 300 K, $X_{\mathrm{CH4}}$ = 0.0 - 1.0. (c) FGM flamelet data at $X_{\mathrm{CH4}}$ = 0.0 and $\phi$ = 1, $T_0$ = 250 - 700 K. (d) FGM flamelet data at $T_0$ = 300 K and $\phi$ = 1, $X_{\mathrm{CH4}}$ = 0.0 - 1.0.}
	\label{Flamelet_data}
\end{figure}

Figure \ref{Flamelet_data} shows the flamelet solutions of the one-dimensional counter-flow diffusion flames (shown as (a) and (b)) and freely-propagating flames (shown as (c) and (d)) in the mixture fraction space and the progress variable space, respectively, at different temperatures and $X$ values. While for the freely-propagating flames, each flamelet has only one mixture fraction value, thus they are shown in condition of equivalence ratio $\phi$ = 1 here. It is found that as the fuel/oxidizer temperature increases, the general temperature increases, indicating that heat loss is important in the flame behavior. As the ammonia fraction in the fuel stream increases, the flame temperature gradually decreases due to the relatively lower heat value of ammonia. For the FGM flamelet data, as the methane fraction in the fuel stream increases, the maximum progress variable and temperature increase.

\section{Numerical setup\label{sec:figtabeqn}} \addvspace{10pt}

\begin{figure}[htbp!]	
	\centering
	\includegraphics[width=192pt]{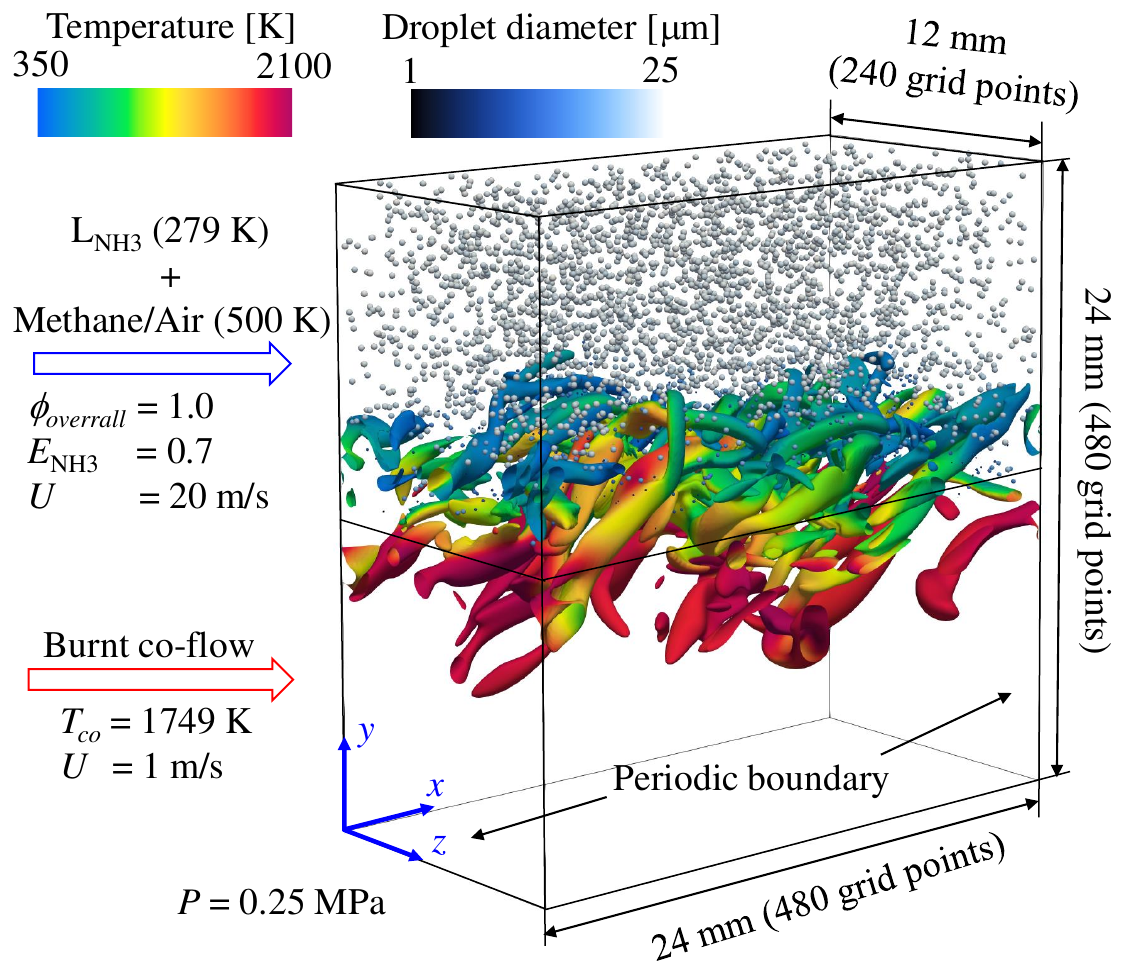}
	\caption{\footnotesize Computational domain and boundary conditions for temporally evolving liquid ammonia flame. The ammonia droplets are colored by diameter. The vortex structure is obtained from the Q criterion (Q = $1 \times 10^7$ $1/s^2$) colored by the local gas temperature. The flow goes from left to right.}
	\label{computational_domain}
\end{figure}

The computational domain of the turbulent mixing layer spans 24 mm in $x$ and $y$ directions, and 12 mm in $z$ direction as shown in Fig. \ref{computational_domain}. A uniform grid resolution of 50 $\mathrm{\mu m}$ is used, resulting in grid points of 480$\times$480$\times$240. The mesh resolution is sufficient to resolve the flame thickness and the Kolmogorov length scale. Liquid ammonia ($L_{\mathrm{NH_3}}$) droplets with a diameter of 25 $\mathrm{\mu m}$ (Note that this is the value after the atomization process) and a premixed mixture of methane/air are injected into the upper half of the computational domain at a velocity of 20 m/s and a carrier gas preheating temperature of 500 K. The heat fraction of ammonia in the total fuel stream is 0.7, the global equivalence ratio is 1.0. All the droplets are randomly distributed in the upper domain. The Stokes number of the droplets is less than 0.6. Then they have good tracking property and the initial droplet velocity is set equal to that of the carrier gas. The initial droplet temperature is 279 K, and the initial ambient pressure is 0.25 MPa. The ammonia droplets undergo the flash evaporation at this condition. The phase change process can be well captured using the combined model \cite{an2023numerical}. This setup is referred to the experimental configurations \cite{okafor2021flame, okafor2021liquid}. For the lower part of the domain, the properties of co-flow are determined by calculated results of $\mathrm{NH_3/CH_4}$/air flames under the same condition. The velocity of the hot co-flow mixture is 1 m/s. The periodic boundary conditions are applied to left/right and front/back sides. The zero-gradient boundary condition is used for the top and bottom sides. To support development of turbulence in the mixing layer, the turbulent flow is generated by the Passot-Pouquet isotropic kinetic spectrum \cite{passot1987numerical} with a fluctuation of 4\% of the jet velocity and an integral length scale of $L_y/6$. The turbulent flow is then superimposed on the main stream.

The PP-DNS and the \textit{a priori} analysis are carried out on the supercomputer Fugaku at the RIKEN Center for Computational Science with in-house solvers, and the computational cost of the PP-DNS is approximately 800,000 core hours by parallel computation with 7200 cores. The Euler framework is employed to solve the governing equations for the gas phase, utilizing the SIMPLE algorithm for calculations. Spatial and transient terms are integrated using a second-order central differencing scheme, while temporal terms are integrated using the Euler-implicit scheme. To maintain numerical stability, the Courant number is constrained to be below 0.1, leading to a time step of approximately 0.4 $\mathrm{\mu s}$.

\section{Results and discussion\label{sec4}} \addvspace{10pt}

\subsection{Analysis of liquid ammonia/methane co-fired flame\label{subsec:subsection4.1}} \addvspace{10pt}

\begin{figure*}[h!]
	\centering
	\includegraphics[width=350pt]{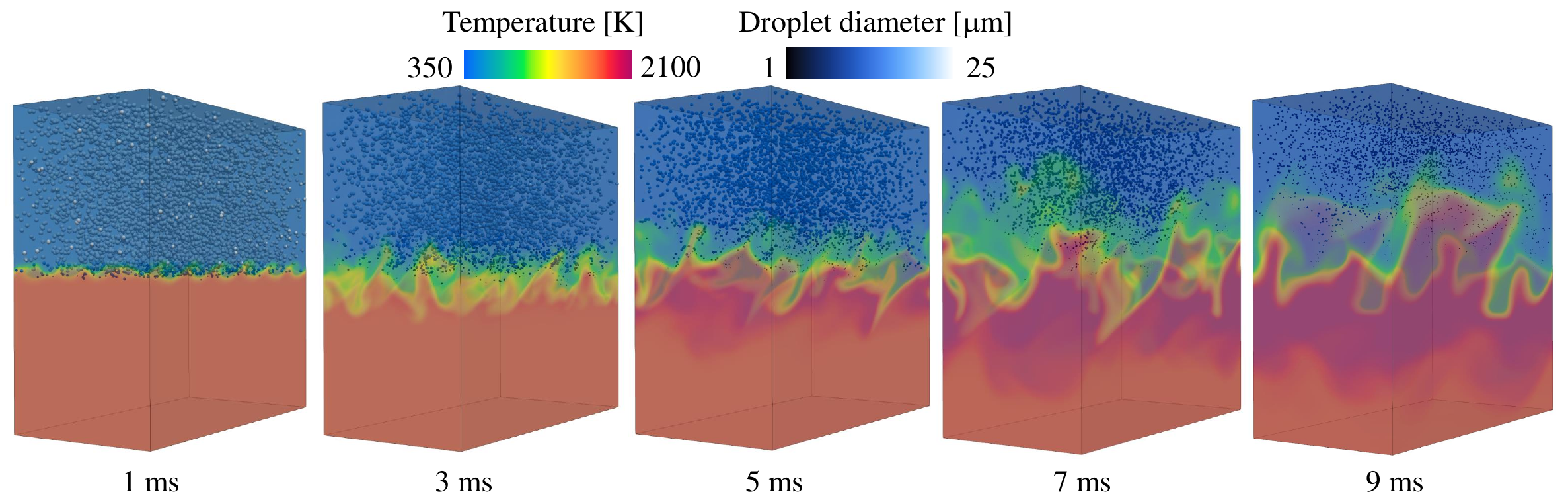}
	\vspace{10 pt}
	\caption{\footnotesize Temporal evolution of gas phase temperature fields and distributions of ammonia droplets for liquid ammonia flame from 1 - 9 ms.}
	\label{temperal_evolution}
\end{figure*}

Figure \ref{temperal_evolution} shows the temporal evolution results of PP-DNS for the liquid ammonia/methane co-fired flame from initial condition to burnt state. The regions represent temperature fields and distributions of ammonia droplets. Initially, the fresh mixture and the burned co-flow are located in the upper and lower parts of the computational domain, respectively. Before the reaction can proceed, the ammonia droplets begin to evaporate. As mixing layer turbulence develops, the evaporated liquid ammonia and carrier gas mix with the hot co-flow. Due to the flash evaporation of the ammonia droplets, the temperature drops briefly before ignition. After 5 ms, the temperature increases significantly and the diameter of the ammonia droplets continues to decrease. As mixing and turbulence further develop, the flame is ignited and propagates upward and downward. Due to flash evaporation, the liquid ammonia evaporates completely before reaching the high temperature region of the flame, making it difficult to see the interaction between the droplets and the flame.

\begin{figure}[h!]
	\centering
	\includegraphics[width=192pt]{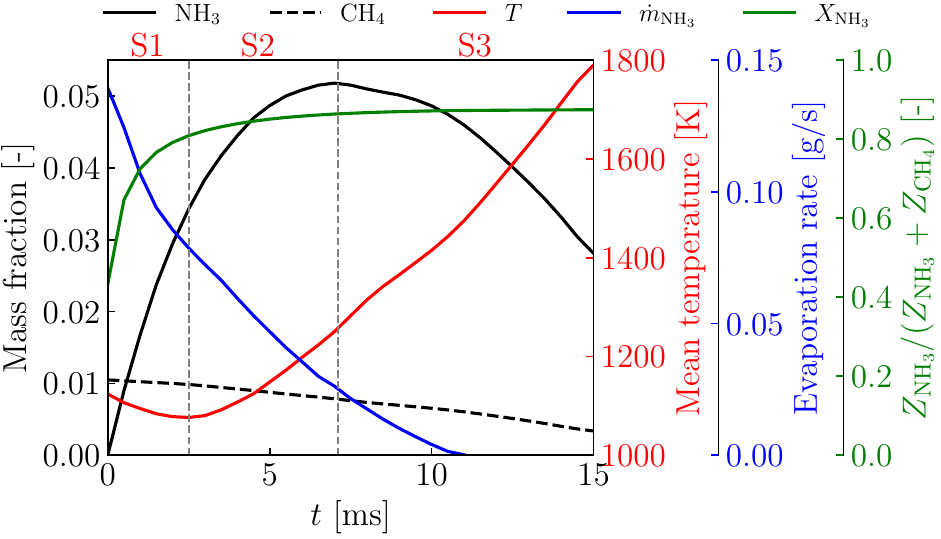}
	\caption{\footnotesize Temporal evolution of the volume-averaged mass fractions of NH$_3$ (black solid line) and CH$_4$ (black dash line), mean temperature of the carrier gas (red line), evaporation rate of ammonia droplets (blue line), and fraction of $Z_{\mathrm{NH_3}}$ in $Z_{\mathrm{NH_3}}+Z_{\mathrm{CH_4}}$ (green line). S1, S2, and S3 represent mixing, methane-dominated, and fully reacting stages, respectively.}
	\label{stages}
\end{figure}

The fuel stream contains ammonia droplets and methane. The phase change process of liquid ammonia will absorb the heat from the surrounding gas, resulting in strong local heat loss. The ignition delay time of methane is much less than that of ammonia. Methane is consumed first, and then ammonia is piloted by ignition of methane and consumed. Therefore, the combustion process of liquid ammonia/methane co-fired flame can be divided into three stages, as shown in Fig. \ref{stages}. In the first stage (S1), the ammonia droplets undergo  rapid flash evaporation and the volume-averaged temperature decreases continuously, indicating an inert evaporation and mixing stage. In this stage, the fraction of ammonia in the total fuel stream increases significantly. In the second stage (S2), liquid ammonia continues to evaporate. The heat release from the reaction is greater than the heat loss caused by the phase change. The temperature then begins to rise. The main contribution of HRR comes from the reaction of methane. This stage is called the methane-dominated stage. In the third stage (S3), the evaporation rate of liquid ammonia begins to be lower than the combustion consumption rate of ammonia, and its mass fraction decreases. In this stage, the ammonia continues to be consumed, and the temperature continues to rise. This stage is called the fully reacting stage. The \textit{a priori} analysis is performed in the Section \ref{subsec:subsection4.2} for these three stages.

\begin{figure}[h!]	
	\centering
	\includegraphics[width=192pt]{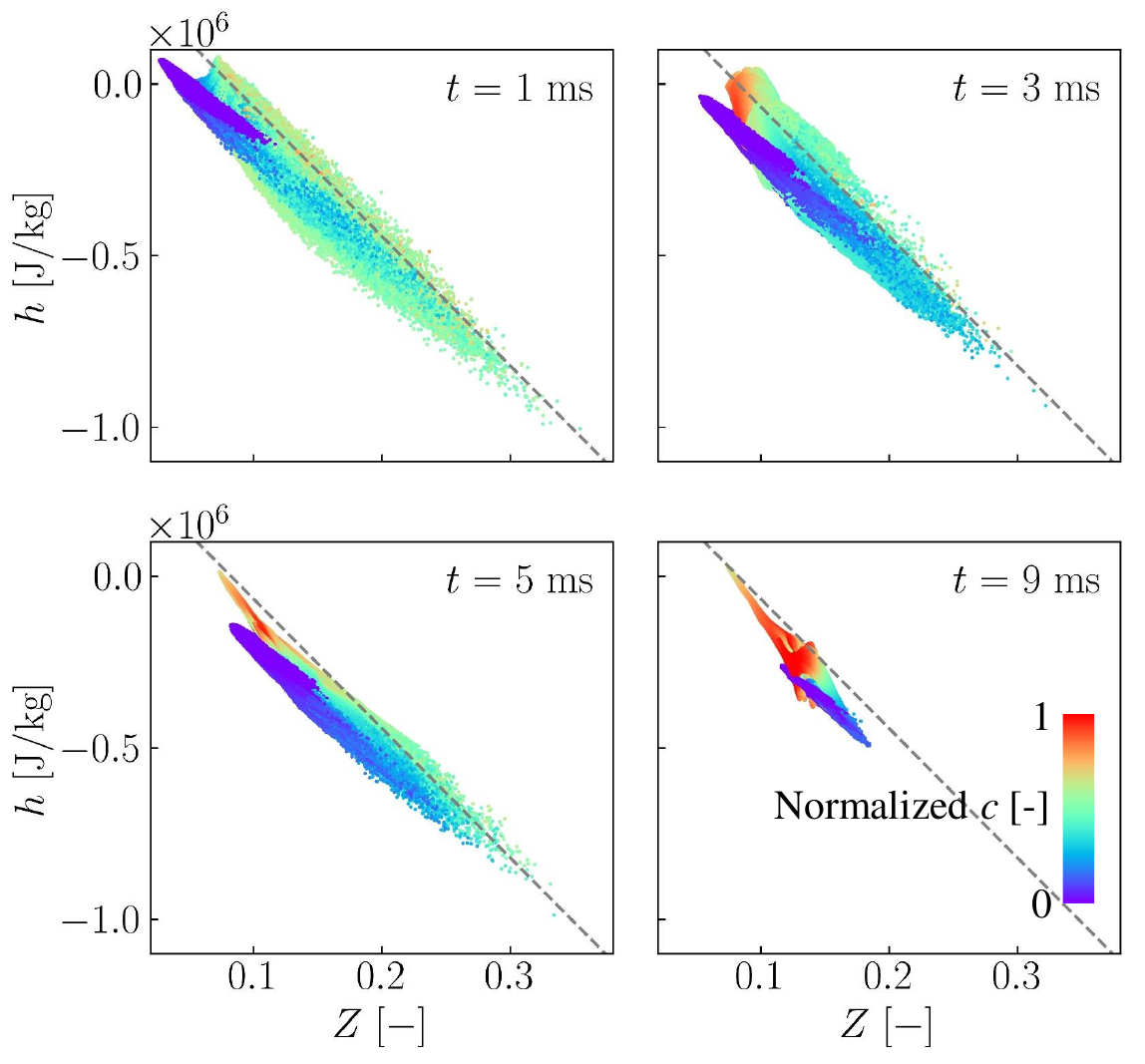}
	\caption{\footnotesize Scatter plots of total enthalpy and mixture fraction from the whole computational domain at different times. The scatters are colored by normalized $c$.}
	\label{Heat_loss}
\end{figure}

Because of the large latent heat of ammonia droplets, the rapid phase change process causes large heat loss. Figure \ref{Heat_loss} shows the relationship between total enthalpy and mixture fraction at different times. The scatters are colored by normalized $c$. The dashed lines represent the flamelet results with $E_{\mathrm{NH_3}}$ = 0.7 and an initial temperature of 500 K. Before 5 ms, the scatters are clearly divided into two clusters, with the cluster with a larger value of $c$ near the dashed line and the cluster with a smaller value of $c$ far below the dashed line, representing the hot co-flow and the fresh mixture stream, respectively. With the development of mixing, the hot co-flow is mixed with unburned gas, the cluster with a smaller value of $c$ decreases, and $c$ increases. It can be seen that at all times, the vast majority of the scatters are below the dashed line, indicating a significant heat loss. Therefore, when simulating a liquid ammonia flame using flamelet-based models, the consideration of enthalpy becomes essential to accurately capture and describe the heat loss phenomenon.

\subsection{A priori analysis of the extended flamelet-based models\label{subsec:subsection4.2}} \addvspace{10pt}

According to the analysis of PP-DNS, there are three stages for liquid ammonia/methane co-fired flame. The \textit{a priori} analysis is carried out to evaluate the performance of the extended FPV, FGM, and hybrid models for these three stages compared with the results of PP-DNS as a benchmark. The four control variables ($c, Z, X, h_{norm}$) are used as the input parameters for the flamelet look-up tables. $c$ is computed from the transported mass fraction. The two mass fractions $Z_{\mathrm{CH_4}}$ and $Z_{\mathrm{NH_3}}$ are transported. Then, $Z$ and $X$ are calculated by $Z = Z_{\mathrm{CH_4}} + Z_{\mathrm{NH_3}}$ and $X = Z_{\mathrm{CH_4}}/\left(Z_{\mathrm{CH_4}} + Z_{\mathrm{NH_3}}\right)$, respectively. $h_{norm}$ is calculated form the transport equation of enthalpy and normalized to 0-1. This analysis provides a direct assessment of the ability of the flamelet-based model to accurately reproduce the reference solution.

\begin{figure}[h!]	
	\centering
	\includegraphics[width=192pt]{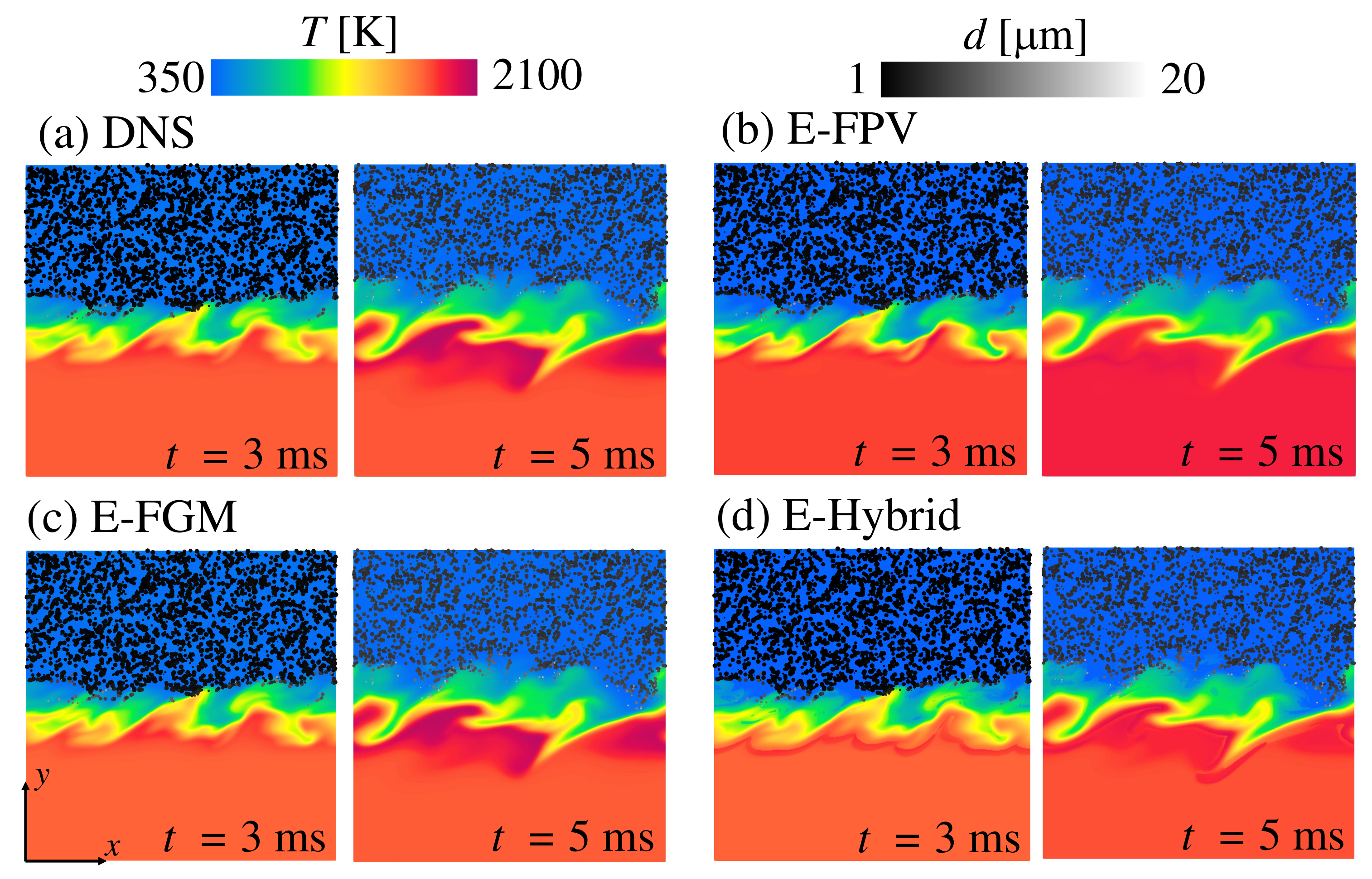}
	\caption{\footnotesize Gas phase temperature and ammonia droplets colored by their diameter in a slice at $z=L_z/2$ at the selected times for the results of (a) DNS, (b) E-FPV, (c) E-FGM, and (d) E-Hybrid models. }
	\label{T_fields}
\end{figure}

Figure \ref{T_fields} shows the temporal evolution of temperature fields at selected times for DNS and flamelet-based models, in a slice at $z = L_z/2$. The DNS results can be regarded as the state-of-the-art validation, as shown in Fig. \ref{T_fields}(a). At $t$ = 3 ms, the temperature has not yet risen. At $t$ = 5 ms, the fresh mixture is ignited, and the temperature of the ignition kernel increases. The temperature of fuel stream decreases slightly due to flash evaporation of the ammonia droplets. The temperature of the hot co-flow does not change much. Figure \ref{T_fields}(b) shows the predictions of the E-FPV model. E-FPV overestimates the temperature of the co-flow and does not predict the temperature non-uniformity of the mixing layer. In particular, the temperature in the reaction zone is not well predicted. Figure \ref{T_fields}(c) shows the prediction of the E-FGM model. It can be found that E-FGM prediction has a good agreement with DNS solution. The flame structure, the temperature stratification, and high temperature zone near the reaction zone are all well predicted. This is mainly due to the rapid flash evaporation effect of liquid ammonia, which causes the droplets to evaporate completely before coming into interact with the flame and has enough time to mix with air, resulting in the reaction mainly taking place under the premixed mode. Therefore, the E-FGM model can predict the results well and its performance far exceeds that of E-FPV. As for the E-Hybrid model, since the predictions of the E-FGM model are already accurate enough under the current conditions, the overall prediction accuracy will be lowered after adding the E-FPV model. There is a certain gap between the prediction of the E-Hybrid model and the benchmark, but the difference is smaller than what the E-FPV model predicts, as shown in Fig. \ref{T_fields}(d).

\begin{figure}[h!]	
	\centering
	\includegraphics[width=192pt]{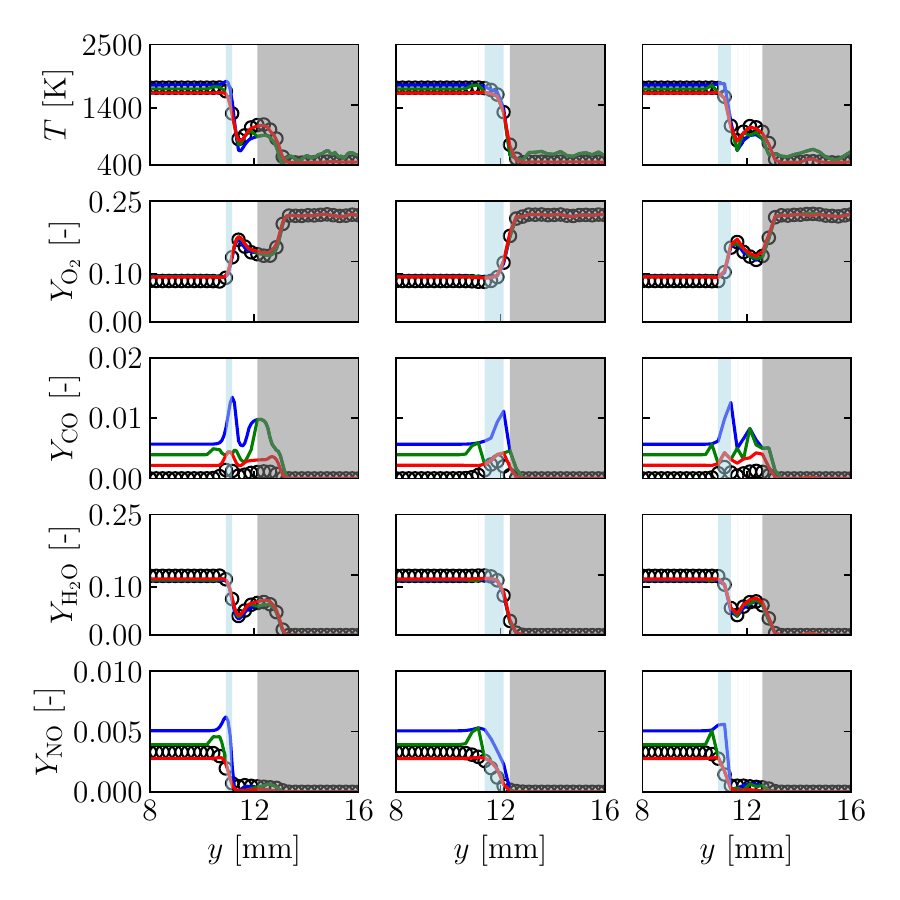}
	\caption{\footnotesize \textit{A priori} analysis along different spatial locations (Left: $x/L_x$ = 1/4, middle: $x/L_x$ = 1/2, right: $x/L_x$ = 3/4) between the results of DNS (black dots) and E-FPV (blue line), E-FGM (red line), E-Hybrid models (green line) at $t$ = 1 ms (S1). The gray and light blue regions represent the non-premixed and premixed combustion modes, respectively.}
	\label{s1}
\end{figure}

Figure \ref{s1} shows the comparisons between the DNS results (black dots) and the \textit{a priori} profiles of the E-FPV (blue line), E-FGM (red line), and E-Hybrid (green line) models at $t$ = 1 ms, which represents the inert mixing stage, along different spatial locations. The gray and light blue regions represent the non-premixed and premixed combustion mode, respectively. In this stage, the turbulence in the mixing layer has not yet developed, as seen in Fig. \ref{temperal_evolution}. Numerous ammonia droplets evaporate and absorb heat from the carrier gas. The ignition kernel is located around $y$ = 12 mm, and will evolve up and down over time. For the temperature field, the predictions of the three models are relatively accurate. It can be found that E-FPV slightly overestimates the value. E-FGM slightly underestimates it on the co-flow side. For $Y_{\mathrm{O_2}}$ and $Y_{\mathrm{H_2O}}$, it can be seen that these three flamelet-based models all have good agreements with DNS. However, the E-FPV model significantly overestimates the profile of CO and NO, especially in the co-flow. This deviation could be attributed to the fact that co-flow is a combustible mixture formed after the premixed gas burns out, and the FPV model based on the diffusion flame surface is not accurate in predicting minor components such as CO and NO. In contrast, the E-FGM model predicts the value of NO very well, which is in good agreement with the results of DNS for all the three spatial locations. The results of the E-Hybrid model are between those of the E-FPV and E-FGM models, and also overestimate CO and NO. In addition, it can be seen that the proportion of the premixed region (light blue) is very small in stage S1.

\begin{figure}[htbp!]	
	\centering
	\includegraphics[width=192pt]{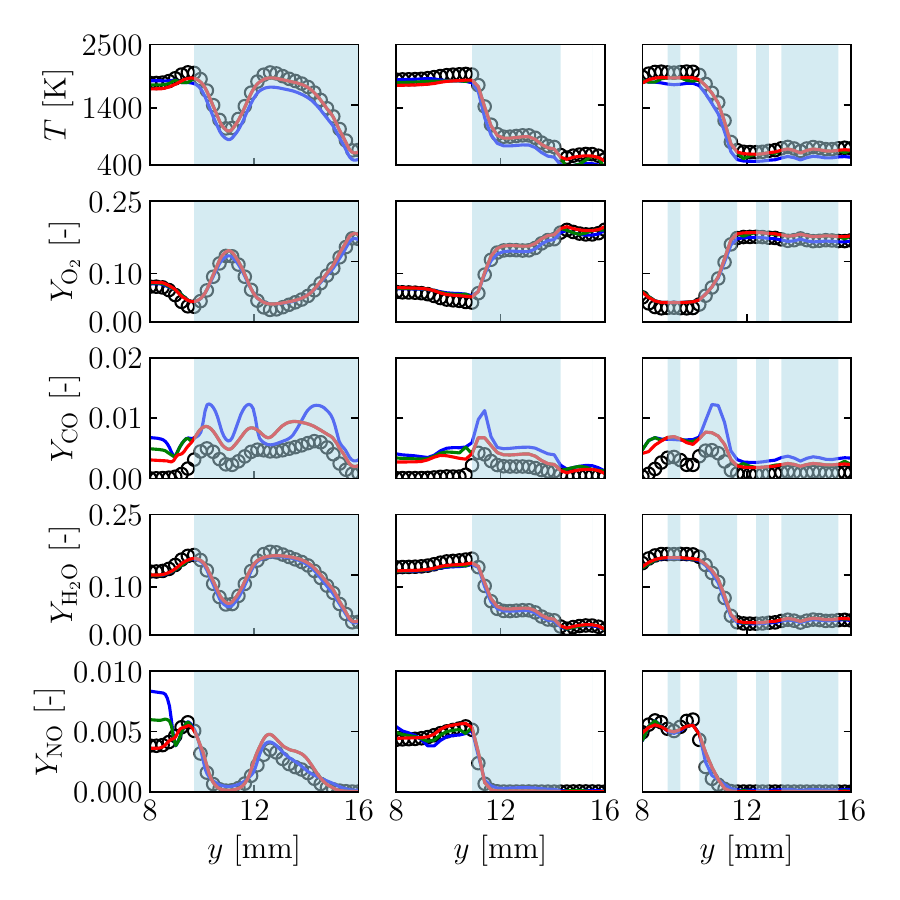}
	\caption{\footnotesize \textit{A priori} analysis along different spatial locations (Left: $x/L_x$ = 1/4, middle: $x/L_x$ = 1/2, right: $x/L_x$ = 3/4) between the results of DNS (black dots) and E-FPV (blue line), E-FGM (red line), E-Hybrid models (green line) at $t$ = 5 ms (S2). The gray and light blue regions represent the non-premixed and premixed combustion modes, respectively.}
	\label{s2}
\end{figure}

Figure \ref{s2} shows the comparisons between the DNS results and the \textit{a priori} profiles of the E-FPV, E-FGM, and E-Hybrid models at $t$ = 5 ms, which represents the methane-dominated stage. In this stage, the ignition kernel propagates towards lower and upper directions, and the temperature increases significantly. Due to the rapid evaporation and mixing process, most of the region features premixed combustion mode. For the temperature field, the E-FPV model underestimates the result. The superiority of the E-FGM and E-Hybrid models retains. For the prediction of the major components, such as O$_2$ and H$_2$O, all three models have a good agreement with DNS. Note that E-FGM predicts intermediate components well, although CO has some overestimation, it is closer to the DNS results compared to other models.

\begin{figure}[htbp!]	
	\centering
	\includegraphics[width=192pt]{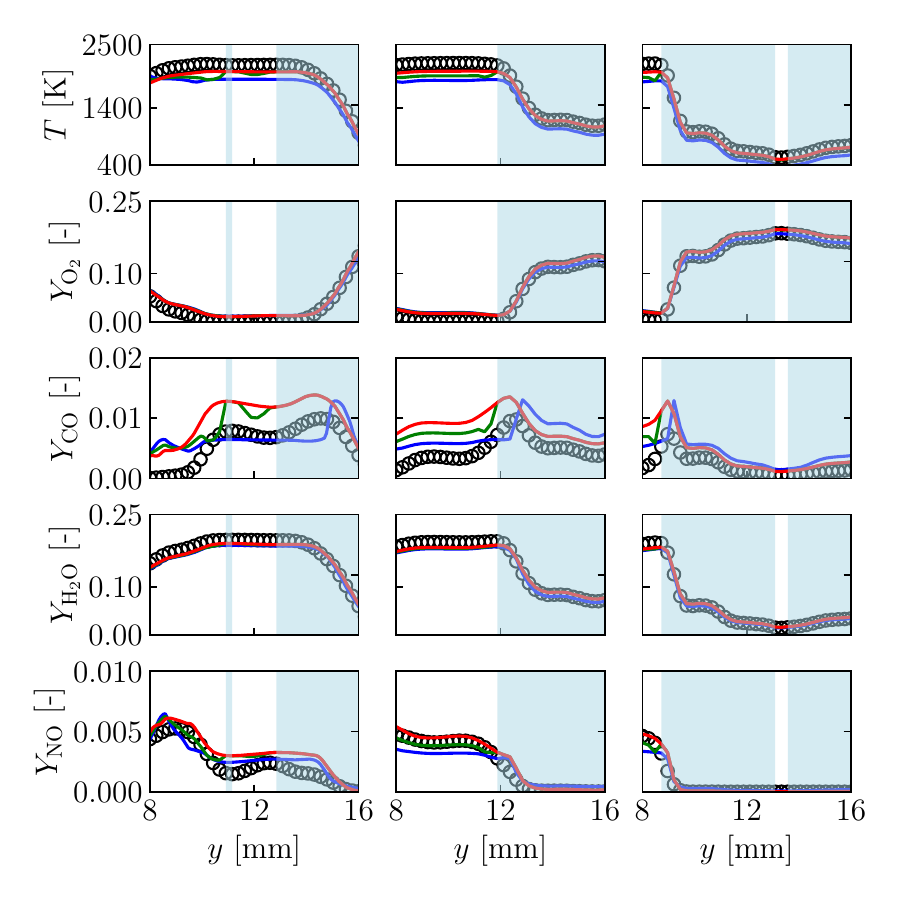}
	\caption{\footnotesize \textit{A priori} analysis along different spatial locations (Left: $x/L_x$ = 1/4, middle: $x/L_x$ = 1/2, right: $x/L_x$ = 3/4) between the results of DNS (black dots) and E-FPV (blue line), E-FGM (red line), E-Hybrid models (green line) at $t$ = 9 ms (S3). The gray and light blue regions represent the non-premixed and premixed combustion modes, respectively.}
	\label{s3}
\end{figure}

Figure \ref{s3}shows the comparisons between the DNS results and the \textit{a priori} profiles of the E-FPV, E-FGM, and E-Hybrid models at $t$ = 9 ms, which represents the fully reacting stage. In this stage, the turbulence is fully developed, and most of the droplets have completely evaporated. The fresh mixture continues to be consumed, and the temperature continues to rise. The mixing proceeds thoroughly, the premixed mode dominates the whole region. Therefore, the results of the E-FGM and E-hybrid models are nearly the same (see red and green lines). The superiority of the E-FGM model retains, especially for the prediction of NO mass fraction. For the gas temperature, E-FGM and E-Hybrid models can also give better predictions. 

Overall, the extended flamelet-based models can give a reasonable prediction for liquid ammonia/methane co-fired flame considering the complex fuel streams and large heat loss. The E-FPV and E-Hybrid models give relatively large differences in temperature and minor products, which is due to the combustion mode being mainly premixed mode. The agreement between the DNS and \textit{a priori} E-FGM results is quite good for all the three stages. The above results indicate the superiority of the E-FGM model in predicting the liquid ammonia/methane co-fired flame.

\section{Conclusions\label{sec:unnum}} \addvspace{10pt}

In the present study, the extended flamelet-based models for the liquid ammonia combustion were developed and evaluated in a temporally evolving mixing layer. Specifically, a PP-DNS was firstly performed, and three reacting stages were identified: mixing stage, methane-dominated stage, and fully reacting stage. It was observed that the effect of heat loss is significant in liquid ammonia combustion. Then, E-FPV, E-FGM, and E-Hybrid models were developed and evaluated through the \textit{a priori} study. By comparing the predictions of the E-FPV, E-FGM, and E-Hybrid models with the DNS solutions as benchmarks, it can be seen that for the simulation of liquid ammonia combustion, the E-FGM model has a better performance than the E-FPV and E-Hybrid models. This can be attributed to the rapid flash evaporation and sufficient mixing of the liquid ammonia. In the future study, the E-FGM model can be used for the numerical simulation of liquid ammonia combustion and liquid ammonia/methane co-fired flame.

\acknowledgement{Declaration of competing interest} \addvspace{10pt}

The authors declare that they have no known competing financial interests or personal relationships that could have appeared to influence the work reported in this paper.

\acknowledgement{Acknowledgments} \addvspace{10pt}

This work was partially supported by MEXT as “Program for Promoting Researches on the Supercomputer Fugaku” (Development of the Smart design system on the supercomputer “Fugaku” in the era of Society 5.0) (JPMXP1020210316). This research used the computational resources of supercomputer Fugaku provided by the RIKEN Center for Computational Science (Project ID: hp230321, hp220141). Z.H.A is grateful to the China Scholarship Council (Project No. 202206280012). J.K.X and R.K especially thank the support from Japan Society for the Promotion of Science (Grant No: 21P20351).


 \footnotesize
 \baselineskip 9pt


\bibliographystyle{pci}
\bibliography{PCI_LaTeX}


\newpage

\small
\baselineskip 10pt



\end{document}